\documentclass[aps,prl,twocolumn,groupedaddress,floatfix]{revtex4-1}
\usepackage{graphicx}
\usepackage{amssymb}
\usepackage{color}
\usepackage{url}

\newcommand{\bea}{\begin{eqnarray}}
\newcommand{\eea}{\end{eqnarray}}
\newcommand{\beas}{\begin{eqnarray*}}
\newcommand{\eeas}{\end{eqnarray*}}
\newcommand{\set}[1]{\left\{ #1 \right\}}

\definecolor{red}{rgb}{1,0,0}
\begin{document}

\title{Rapid and deterministic estimation of probability densities\\ using scale-free field theories}

\author{Justin B. Kinney}
\email[Please direct correspondence to ]{jkinney@cshl.edu}

\affiliation{Simons Center for Quantitative Biology, Cold Spring Harbor Laboratory, Cold Spring Harbor, NY, 11724, USA}

\date{18 April 2014}

\begin{abstract}
The question of how best to estimate a continuous probability density from finite data is an intriguing open problem at the interface of statistics and physics. Previous work has argued that this problem can be addressed in a natural way using methods from statistical field theory. Here I describe new results that allow this field-theoretic approach to be rapidly and deterministically computed in low dimensions, making it practical for use in day-to-day data analysis. Importantly, this approach does not impose a privileged length scale for smoothness of the inferred probability density, but rather learns a natural length scale from the data due to the tradeoff between goodness-of-fit and an Occam factor. Open source software implementing this method in one and two dimensions is provided.
\end{abstract} 


\maketitle

Suppose we are given $N$ data points, $x_1, x_2, \ldots, x_N$, each of which is a $D$-dimensional vector drawn from a smooth probability density $Q_{true}(x)$. How might we estimate $Q_{true}$ from these data? This classic statistics problem is known as ``density estimation'' \cite{Eggermont:2001} and is routinely encountered in nearly all fields of science. Ideally, one would first specify a Bayesian prior $p(Q)$ that weights each density $Q(x)$ according to some sensible measure of smoothness. One would then compute a Bayesian posterior $p(Q | {\rm data})$ identifying which densities are most consistent with both the data and the prior. However, a practical implementation of this straight-forward approach has yet to be developed, even in low dimensions. 

This paper discusses one such strategy, the main theoretical aspects of which were worked out by Bialek et al.\ in 1996 \cite{Bialek:1996p292}.  One first assumes a specific smoothness length scale $\ell$. A prior $p(Q | \ell)$ that strongly penalizes fluctuations in $Q$ below this length scale is then formulated in terms of a scalar field theory. The maximum \emph{a posteriori} (MAP) density $Q_\ell$, which maximizes $p(Q | \ell, {\rm data})$ and serves as an estimate of $Q_{true}$, is then computed as the solution to a nonlinear differential equation. This approach has been implemented and further elaborated by others \cite{Nemenman:2002p152, Holy:1997p1125, Periwal:1997p1128, Aida:1999p1130, Schmidt:2000p1132, Lemm:2003, Nemenman:2005, Ensslin:2009}; a connection to previous literature on ``maximum penalized likelihood'' \cite{Eggermont:2001} should also be noted.

Left lingering is the question of how to choose the length scale $\ell$. Bialek et al.\ argued, however, that the data themselves will typically select a natural length scale due to the balancing effects of goodness-of-fit (i.e.\ the posterior probability of $Q_\ell$) and an Occam factor (reflecting the entropy of model space \cite{Balasubramanian:1997p1127}).  Specifically, if one adopts a ``scale-free'' prior $p(Q)$, defined as a linear combination of scale-dependent priors $p(Q | \ell)$, then the posterior distribution over length scales, $p(\ell | {\rm data})$, will become sharply peaked in the large data limit. This important insight was confirmed computationally by Nemenman and Bialek \cite{Nemenman:2002p152} and provides a compelling alternative to cross-validation, the standard method of selecting length scales in statistical smoothing problems \cite{Eggermont:2001}.

However computing $p(\ell | {\rm data})$ requires first computing $Q_\ell$ at every relevant length scale, i.e.\ solving an infinite compendium of nonlinear differential equations. Nemenman and Bialek \cite{Nemenman:2002p152} approached this problem by computing $Q_\ell$ at a finite, pre-selected set of length scales. Although this strategy yielded important results, it also has significant limitations. First, it is unclear how to choose the set of length scales needed to estimate $Q_{true}$ to a specified accuracy. Second, as was noted \cite{Nemenman:2002p152}, this strategy is very computationally demanding. Indeed, no implementation of this approach has since been developed for general use, and performance comparisons to more standard density estimation methods have yet to be reported. 

Here I describe a rapid and deterministic homotopy method for computing $Q_\ell$ to a specified accuracy at all relevant length scales. This makes low-dimensional density estimation using scale-free field-theoretic priors practical for use in day-to-day data analysis. The open source ``Density Estimation using Field Theory'' (DEFT) software package, available at \url{github.com/jbkinney/13_deft}, provides a Python implementation of this algorithm for 1D and 2D problems. Simulation tests show favorable performance relative to standard Gaussian mixture model (GMM) and kernel density estimation (KDE) approaches \cite{Eggermont:2001}.

Following \cite{Bialek:1996p292, Nemenman:2002p152} we begin by defining $p(Q)$ as a linear combination of scale-dependent priors $p(Q | \ell)$:
\bea
p(Q) = \int_0^{\infty} d\ell\ p(Q | \ell)\ p(\ell). \label{eq:mixture_in_Q}
\eea
Adopting the Jeffreys prior $p(\ell) \sim \ell^{-1}$ renders $p(Q)$ covariant under a rescaling of $x$ \cite{Balasubramanian:1997p1127}. Our ultimate goal will be to compute the resulting posterior,
\bea
p(Q | {\rm data}) &=& \int_0^{\infty} d\ell\ p(Q | \ell, {\rm data})\ p(\ell | {\rm data}). \label{eq:posterior_in_Q}
\eea
As in \cite{Nemenman:2002p152}, we limit our attention to a $D$-dimensional cube having volume $V = L^D$. We further assume periodic boundary conditions on $Q$, and impose $G^D$ grid points ($G$ in each dimension) at which $Q$ will be computed. 

%
%
\begin{figure}[t] 
 \includegraphics{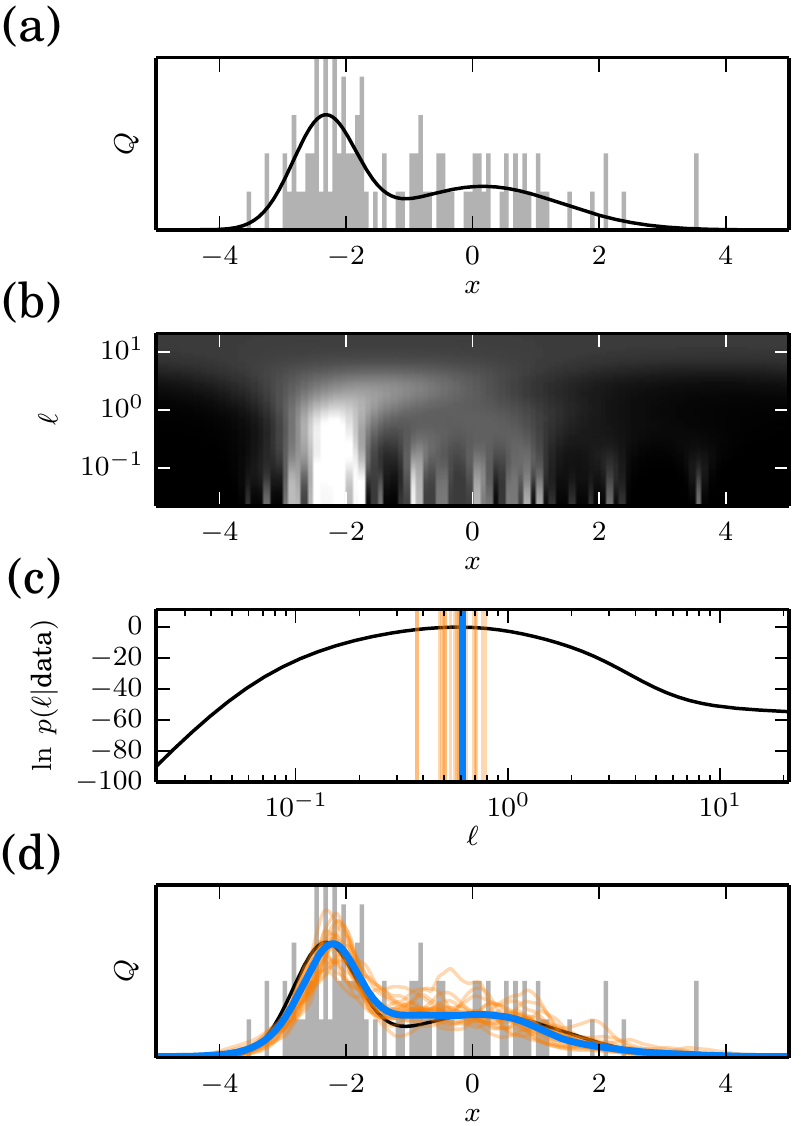}
 \caption{\label{fig1} (Color) Illustration DEFT in one dimension. (a) An example density $Q_{true}(x)$ (black) along with a normalized histogram of $N=100$ sampled data points (gray). (b) Heat map showing all of the MAP densities $Q_{\ell}(x)$ computed using $\alpha = 2$, $G = 100$, and the data from panel (a); lighter shading corresponds to higher probability. (c) Log posterior probability for each length scale $\ell$ shown in panel (b); the y-axis is shifted so that $\ln p(\ell^* | {\rm data}) = 0$. (d) The MAP density $Q^*(x)$ (blue) along with 20 densities (orange) sampled from $p(Q | {\rm data})$ using Eq.\ \ref{eq:sampled_phis}. Length scales $\ell$ corresponding to the MAP and sampled densities are shown in panel (c).}
\end{figure}

To guarantee that each density is positive and normalized, we define $Q$ in terms of a real scalar field $\phi$ as
\bea
Q(x) &=& \frac{e^{-\phi(x)}}{\int d^D x'\ e^{-\phi(x')}}. \label{eq:density_field_correspondence}
\eea
Each $Q$ corresponds to multiple different $\phi$, but there is a one-to-one correspondence with fields $\phi_{nc}$ that have no constant Fourier component. Using this fact, we adopt the standard path integral measure $\mathcal{D} \phi_{nc}$ as the measure on $Q$-space, and define the prior $p(Q | \ell)$ in terms of a field theory on $\phi_{nc}$. In this paper we consider the specific class of priors discussed by \cite{Bialek:1996p292}, i.e.
\bea
p(\phi_{nc} | \ell) = \frac{1}{Z^0_\ell} \exp \left[ - \int d^D x \frac{\ell^{2 \alpha - D}}{2} \phi_{nc} \Delta \phi_{nc} \right]. \label{eq:scale_dependent_prior}
\eea
Here we take $\alpha$ to be a positive integer and define the differential operator $\Delta = (- \nabla^2)^{\alpha}$. $Z^0_{\ell}$ is the corresponding normalization factor. This prior effectively constrains the $\alpha$-order derivatives of $\phi_{nc}$, strongly dampening Fourier modes that have wavelength much less than $\ell$.

Applying Bayes's rule to this prior yields the following exact expression for the posterior \footnote{The identity $a^{-N} = \frac{1}{\Gamma(N)} \int_{-\infty}^{\infty} du \exp \left[ -N u - a e^{-u} \right]$ is used here with $a = (N/V) \int d^D x\ e^{-\phi_{nc}(x)}$ and $u = \phi_c$}:
\bea
p(\phi_{nc} | \ell, {\rm data}) &=& \frac{1}{Z^N_{\ell}} \int_{-\infty}^{\infty} d\phi_c\ e^{- \beta S^N_\ell[\phi]}, \label{eq:scale_dependent_posterior}
\eea
where
\bea
S^N_\ell[\phi] &=& \int d^D x\, \left[ \frac{\phi \Delta \phi}{2}  +  e^t R \phi + e^t \frac{e^{-\phi}}{V} \right] \label{eq:action}
\eea
is the ``action'' described by \cite{Bialek:1996p292} and explored in later work \cite{Lemm:2003, Nemenman:2002p152, Nemenman:2005, Ensslin:2009}. Here, $R(x) = N^{-1} \sum_{n=1}^N \delta(x - x_n)$ is the raw data density, $\phi(x) = \phi_{nc}(x) + \phi_c$,  $t = \ln(N / \ell^{2 \alpha - D})$, $\beta = \ell^{2 \alpha - D}$, and $Z^N_\ell = Z^0_\ell\ \Gamma(N) (V/N)^N p({\rm data} | \ell)$. 

It should be noted that \cite{Bialek:1996p292} used $Q(x) = const \times e^{-\phi(x)}$ in place of Eq.\ \ref{eq:density_field_correspondence}, and enforced normalization using a delta function factor in the prior $p(Q|\ell)$; Eq.\ \ref{eq:action} was then derived using a large $N$ saddle point approximation. The alternative formulation in Eqs.\ \ref{eq:density_field_correspondence} and \ref{eq:scale_dependent_prior} renders the action in Eq.\ \ref{eq:action} exact. 

The MAP density $Q_{\ell}$ corresponds to the classical path $\phi_{\ell}$, i.e.\ the field that minimizes $S^N_{\ell}$. Setting $\delta S_{\ell}^N / \delta \phi = 0$ gives the nonlinear differential equation,
\bea
\Delta \phi_\ell + e^t \left[ R - Q_{\ell} \right] = 0, \label{eq:eom}
\eea
where $Q_{\ell}(x) = e^{-\phi_{\ell}(x)} / V$ is the probability density corresponding to $\phi_{\ell}$. 

The central finding of this paper is that, instead of computationally solving Eq.\ \ref{eq:eom} at select length scales $\ell$, we can compute $\phi_\ell$ at all length scales of interest using a convex homotopy method \cite{Allgower:1990}. First we differentiate Eq.\ \ref{eq:eom} with respect to $t$, yielding
\bea
[e^{-t} \Delta + Q_\ell] \frac{d \phi_\ell}{dt} = Q_\ell - R\label{eq:flow}.
\eea
If we know $\phi_{\ell}$ at any specific length scale $\ell_i$, we can determine $\phi_{\ell}$ at any other length scale $\ell_f$ -- and at all length scales in between -- by integrating Eq.\ \ref{eq:flow} from $\ell_i$ to $\ell_f$. Because $S^N_{\ell}[\phi]$ is a strictly convex function of $\phi$, each $\phi_{\ell}$ so identified will uniquely minimize this action. Moreover, because Eq.\ \ref{eq:action} is exact, each corresponding $Q_\ell$ will fit the data optimally even when $N$ is small. And since the matrix representation of $e^{-t} \Delta + Q_\ell$ is sparse, $d \phi_{\ell} / dt$ can be rapidly computed at each successive value of $t$ using standard sparse matrix methods.
 
To identify a length scale $\ell_i$ from which to initiate integration of Eq.\ \ref{eq:flow}, we look to the large length scale limit, where a weak-field approximation can be used to compute $\phi_{\ell_i}$. Linearizing Eq.\ \ref{eq:eom} and solving for $\phi_\ell$ gives, for $|k| > 0$, $\hat{\phi}_\ell(k) =  - \frac{ V \hat{R}(k)}{1 + \exp[\tau_k - t]}$, where hats denote Fourier transforms, $k \in \mathbb{Z}^D$ indexes the  Fourier modes of the volume $V$, and each $\tau_k = \ln [(2 \pi |k|)^{2 \alpha} L^{D - 2 \alpha}]$ is a log eigenvalue of $V \Delta$. To guarantee that none of the Fourier modes of $\phi_{\ell_i}$ are saturated, $\ell_i$ should correspond to a value $t_i$ that is sufficiently less than $\min_{|k|>0} \tau_k$, i.e.\ $\ell_i \gg N^{\frac{1}{2 \alpha - D}}L$. 

Similarly, we terminate the integration of Eq.\ \ref{eq:flow} at a length scale $\ell_f$ below which Nyquist modes saturate. This yields the criterion $\ell_f \ll n^{\frac{1}{2 \alpha - D}} h$ where $h = L/G$ is the grid spacing and $n = N/G^D$ is the number of data points per voxel.

\begin{figure}[t] 
 \includegraphics{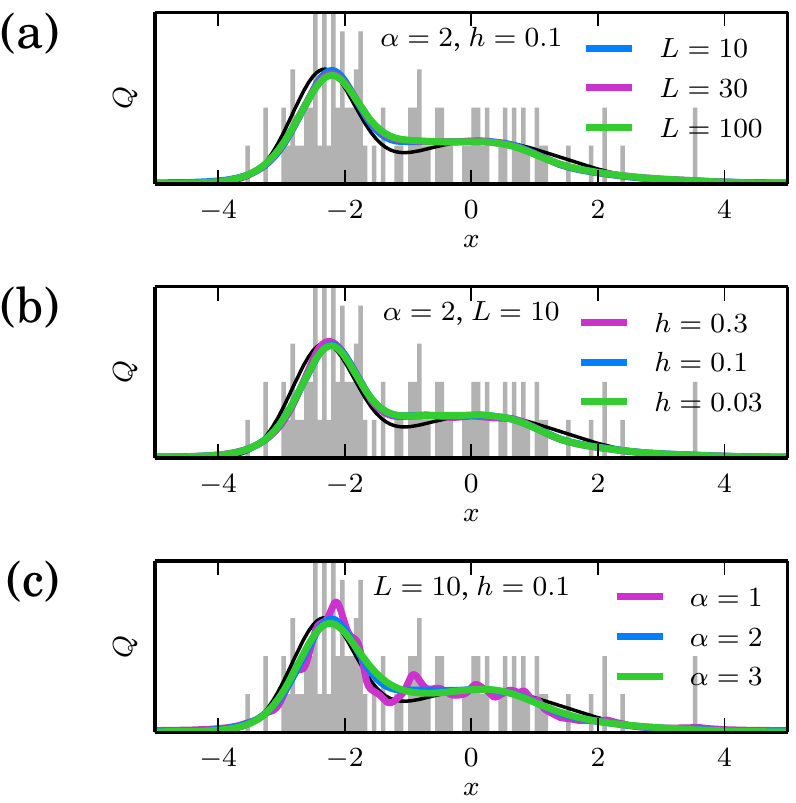}\label{fig2}
 \caption{(Color) Robustness of DEFT to changes in runtime parameters. $Q^*(x)$ was computed using the data from Fig.\ 1 and various choices for (a) the length $L$ of the bounding box, (b) the grid spacing $h$, and (c) the order $\alpha$ of the derivative constrained by the field theory prior. $L = 10$ corresponds to the bounding box shown, and $h = 0.1$ is the grid spacing used for the histogram (gray). $Q_{true}(x)$ is shown in black.}
\end{figure}

\begin{figure}[t] 
 \includegraphics{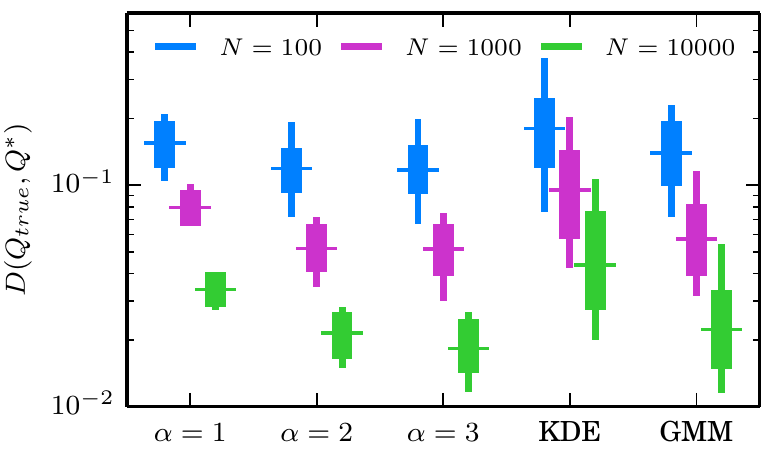} \label{fig3}
 \caption{(Color) Comparison of density estimation methods in one dimension. $10^2$  densities $Q_{true}(x)$ were generated randomly, each as the sum of five Gaussians. Data sets of various size $N$ were then drawn from each $Q_{true}$, after which estimates $Q^*$ were computed using DEFT ($G = 100$, various $\alpha$), KDE (using Scott's rule to set kernel bandwidth), and GMM (using the Bayesian information criterion to choose the number of components). Accuracy was quantified using the geodesic distance $D_{geo}(Q_{true}, Q^*)$ shown in Eq.\ \ref{eq:geo}. Box plots indicate the median, interquartile range, and 5\%-95\% quantile range of these $D_{geo}$ values.}
\end{figure}

Having computed $\phi_\ell$ at every relevant length scale, a semiclassical approximation gives 
\bea
p(\ell | {\rm data}) = const \times \frac{e^{-\beta S^N_\ell [\phi_{\ell}]}}{\sqrt{\beta \det [\Delta + e^t Q_{\ell}]}} \times p(\ell).~~~ \label{eq:semiclassical_approx}
\eea
The MAP probability, $p(\phi_\ell | \ell, {\rm data})$, is represented by the exponential term. This is the ``goodness-of-fit'', which steadily increases as $\ell$ gets smaller. This is multiplied by an Occam factor (the inverse rooted quantity), which steadily decreases as $\ell$ get smaller due to the increasing entropy of model space. As discussed by \cite{Bialek:1996p292, Nemenman:2002p152}, this tradeoff causes $p(\ell | {\rm data})$ to peak at a nontrivial, data-determined length scale $\ell^*$.

The length scale prior $p(\ell)$ must decay faster than $\ell^{-1}$ in the infrared in order for $p(\ell | {\rm data})$ to be normalizable. The need for such regularization reflects redundancy among the priors $p(\phi_{nc} | \ell)$ for $\ell > \ell_i$ that results from the volume $V$ supporting only a limited number of long wavelength Fourier modes. Similar concerns hold in the ultraviolet due to our use of a grid. We therefore set $p(\ell) = 0$ for $\ell > \ell_i$ and $\ell < \ell_f$. 

The posterior density $p(Q | {\rm data})$ can be sampled by first choosing $\ell \sim p(\ell | {\rm data})$, then selecting $\phi \sim p(\phi | \ell, {\rm data})$. This latter step simplifies when $p(\ell | {\rm data})$ is strongly peaked about a specific $\ell^*$ because, in this case, the eigenvalues and eigenfunctions of $\Delta + e^t Q_{\ell}$ do not depend strongly on which $\ell$ is selected and therefore need to be computed only once. $p(\phi | \ell, {\rm data})$ can then be sampled by choosing
\bea
\phi(x) &=& \phi_{\ell}(x) + \sum_{j = 1}^{G^D} \frac{ \eta_j }{\sqrt{\beta \lambda_j}} \psi_j(x), \label{eq:sampled_phis}
\eea
where each $\psi_j$ is an eigenfunction of the operator $\Delta + e^{t^*} Q^*$ satisfying $\int d^D x\ \psi_j(x)^2 = 1$, $\lambda_j$ is the corresponding eigenvalue, and each $\eta_j$ is a normally distributed random variable.

Fig.\ 1 illustrates key steps of the DEFT algorithm. First, the user specifies a data set $\set{x_n}_{n=1}^N$, a bounding box for the data, and the number of grid points to be used. A histogram of the data is then computed using bins that are centered on each grid point (Fig.\ 1a). Next, length scales $\ell_i$ and $\ell_f$ are chosen. Eq.\ \ref{eq:flow} is then integrated to yield $\phi_\ell$ at a set of  length scales between $\ell_i$ and $\ell_f$ chosen automatically by the ODE solver to achieve the desired accuracy. Eq.\ \ref{eq:semiclassical_approx} is then used to compute $p(\ell | {\rm data})$ at each of these length scales, after which $\ell^*$ is identified. Finally, a specified number of densities are sampled from $p(Q | {\rm data})$ using Eq.\ \ref{eq:sampled_phis}. 

DEFT is not completely scale-free because both the box size $L$ and grid spacing $h$ are pre-specified by the user. In practice, however, $Q^*$ appears to be very insensitive to the specific values of $L$ and $h$ as long as the data lie well within the bounding box and the grid spacing is much smaller than the inherent features of $Q_{true}$; see Figs.\ 2a and 2b.

\begin{figure}[t]
\includegraphics{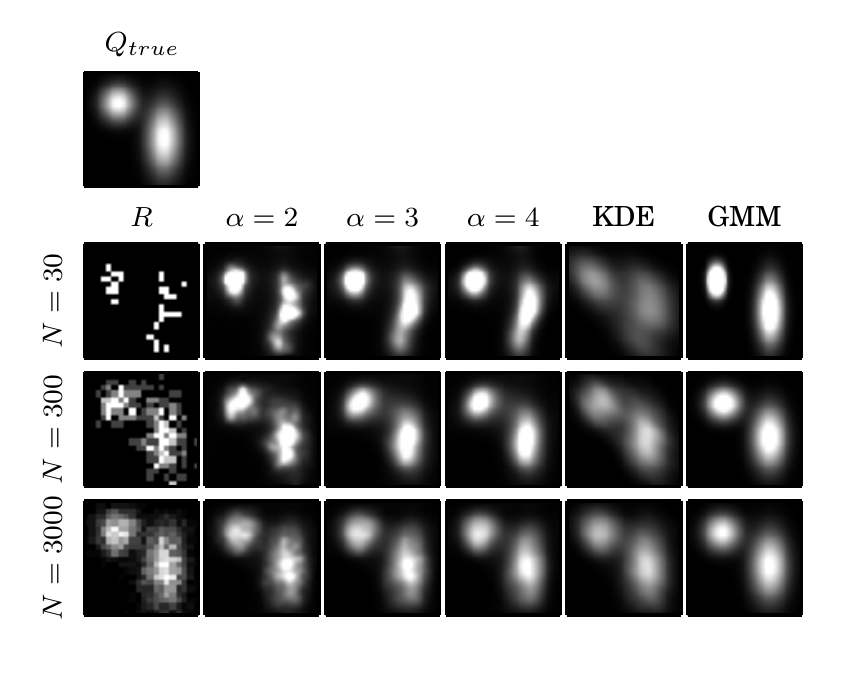} \label{fig4}
\caption{Example density estimates in two dimensions. Shown is a simulated density $Q_{true}$ composed of two Gaussians, data sets of various size $N$ displayed using the raw data density $R$, and the resulting density estimates $Q^*$ computed using DEFT ($G=20$), KDE, or GMM as in Fig.\ 3. The grayscale in all plots is calibrated to $Q_{true}$.}
\end{figure}

It is interesting to consider how the choice of $\alpha$ affects $Q^*$. As Bialek et al.\ have discussed \cite{Bialek:1996p292}, this field-theoretic approach produces ultraviolet divergences in $\phi_{\ell}$ when $\alpha < D/2$. Above this threshold, increasing $\alpha$ typically increases the smoothness of $Q^*$, although not necessarily by much (see Fig.\ 2c). However, larger values of $\alpha$ may necessitate more data before the principal Fourier modes of $Q_{true}$ appear in $Q^*$. Increasing $\alpha$ also reduces the sparseness of the $\Delta$ matrix, thereby increasing the computational cost of the homotopy method. 

To assess how well DEFT performs in comparison to more standard density estimation methods, a large number of data sets were simulated, after which the accuracy of $Q^*$ produced by various estimators was computed. Specifically, the ``closeness'' of $Q_{true}$ to each estimate $Q^*$ was quantified using the natural geodesic distance \cite{Skilling:2007p1126},
\bea
D_{geo}(Q_{true}, Q^*) = 2 \cos^{-1} \left[ \int d^D x \sqrt{Q_{true}Q^*} \right]. \label{eq:geo}
\eea
As shown in Fig.\ 3, DEFT performed substantially better when $\alpha =2$ or 3 than when $\alpha = 1$. This likely reflects the smoothness of the simulated $Q_{true}$ densities. DEFT outperformed the KDE method tested here and, for $\alpha =2$ or 3, performed as well or better than GMM. This latter observation suggests nearly optimal performance by DEFT, since each simulated $Q_{true}$ was indeed a mixture of Gaussians.

In two dimensions, DEFT shows a remarkable ability to discern structure from a limited amount of data (Fig.\ 4). As in 1D, larger values of $\alpha$ give a smoother $Q^*$. However, DEFT requires substantially more computational power in 2D than in 1D due to the increase in the number of grid points and the decreased sparsity of the $\Delta$ matrix. For instance, the computation shown in Fig.\ 1 took about 0.3 sec, while the DEFT computations shown in Fig.\ 4 took about 1-3 sec each \footnote{Computation times were assessed on a computer having a 2.8 GHz dual core processor, 16 GB of RAM, and running the Canopy Python distribution (Enthought).}. 

Field-theoretic density estimation faces two significant challenges in higher dimensions. First, the computational approach described here is impractical for $D \gtrsim 3$ due to the enormous number of grid points that would be needed. It should be noted, however, that the 1D field theory discussed by Holy \cite{Holy:1997p1125} allows $Q_{\ell}$ to be computed without using a grid. It may be possible to extend this approach to higher dimensions. 

The ``curse of dimensionality'' presents a more fundamental problem. As discussed by Bialek et al.\ \cite{Bialek:1996p292}, this manifests in the fact that increasing $D$ requires a proportional increasing in $\alpha$, i.e.\ in one's notion of ``smoothness.'' This likely indicates a fundamental problem with using $\Delta = (- \nabla^2)^{\alpha}$ to define high dimensional priors. Using a different operator for $\Delta$, e.g.\ one with reduced rotational symmetry, might provide a way forward. 

I thank Gurinder Atwal, Anne-Florence Bitbol, Daniel Ferrante, Daniel Jones, Bud Mishra, Swagatam Mukhopadhyay, and Bruce Stillman for helpful conversations. Support for this work was provided by the Simons Center for Quantitative Biology at Cold Spring Harbor Laboratory. 

\bibliography{deft_revised}

\end{document}